\documentclass[9pt,sigconf,authorversion]{acmart}
\usepackage[utf8]{inputenc}
\usepackage[T1]{fontenc}
\usepackage[nohyperlinks]{acronym}
\usepackage[binary-units, per-mode=symbol, detect-weight=true, detect-family=true, group-digits=integer, group-separator={{,}}, separate-uncertainty]{siunitx}
\usepackage[abbreviations, xlatin]{foreign}
\usepackage{booktabs}
\usepackage{csquotes}
\usepackage{xspace}
\usepackage{subfig}
\usepackage[inline]{enumitem}
\usepackage{xcolor}
\usepackage{pifont}
\usepackage{ifthen}
\usepackage{hyperref}
\usepackage{circledsteps}

\newcommand{\vn}[1]{\ensuremath{\boldsymbol{#1}}}

\newcommand{\sys}{\textsc{ZipLine}\xspace}
\title{ZipLine: In-Network Compression at Line Speed}

\author{Sébastien Vaucher}
\affiliation{\institution{Université de Neuchâtel}
	\city{Neuchâtel}
	\country{Switzerland}
}
\email{sebastien.vaucher@unine.ch}

\author{Niloofar Yazdani}
\affiliation{\institution{Aarhus Universitet}
	\city{Aarhus}
	\country{Denmark}
}
\email{n.yazdani@eng.au.dk}

\author{Pascal Felber}
\affiliation{\institution{Université de Neuchâtel}
	\city{Neuchâtel}
	\country{Switzerland}
}
\email{pascal.felber@unine.ch}

\author{Daniel E. Lucani}
\affiliation{\institution{Aarhus Universitet}
	\city{Aarhus}
	\country{Denmark}
}
\email{daniel.lucani@eng.au.dk}

\author{Valerio Schiavoni}
\affiliation{\institution{Université de Neuchâtel}
	\city{Neuchâtel}
	\country{Switzerland}
}
\email{valerio.schiavoni@unine.ch}

\acmYear{2020}\copyrightyear{2020}
\setcopyright{acmlicensed}
\acmConference[CoNEXT '20]{The 16th International Conference on emerging Networking EXperiments and Technologies}{December 1--4, 2020}{Barcelona, Spain}
\acmBooktitle{The 16th International Conference on emerging Networking EXperiments and Technologies (CoNEXT '20), December 1--4, 2020, Barcelona, Spain}
\acmPrice{15.00}
\acmDOI{10.1145/3386367.3431302}
\acmISBN{978-1-4503-7948-9/20/12}
\setlist[enumerate,1]{itemjoin={{; }}, itemjoin*={{, and }}}
\pgfkeys{/csteps/inner color=white}
\pgfkeys{/csteps/outer color=black}
\pgfkeys{/csteps/fill color=black}
\pgfkeys{/csteps/inner ysep=2pt}
\pgfkeys{/csteps/inner xsep=2pt}

\newcommand{\pfoursixteen}{$\text{P}4_{16}$\xspace}

\DeclareSIUnit{\packet}{pkt}

\acrodef{GD}{generalized deduplication}

\acrodef{TNA}{Tofino native architecture}
\acrodef{SDE}{software development environment}
\acrodef{BfRt}{Barefoot runtime}

\acrodef{TTL}{time to live}
\acrodefplural{TTL}{times to live}
\acrodef{LRU}{least recently used}
\acrodefindefinite{LRU}{an}{a}
\acrodef{CRC}{cyclic redundancy check}
\acrodef{TTL}{time to live}
\acrodef{MSB}{most significant bit}
\acrodefindefinite{MSB}{an}{a}
\acrodef{RTT}{round-trip time}
\acrodefindefinite{RTT}{an}{a}
\acrodef{ASIC}{application-specific integrated circuit}
\acrodef{EC}{erasuring coding}
\acrodef{MTU}{maximum transmission unit}
\acrodef{FPGA}{field-programmable gate array}
\acrodef{IoT}{Internet of Things} 


\begin{document}
	\begin{abstract}
		Network appliances continue to offer novel opportunities to offload processing from computing nodes directly into the data plane.
One popular concern of network operators and their customers is to move data increasingly faster. 
A common technique to increase data throughput is to compress it before its transmission.
However, this requires compression of the data---a time and energy demanding pre-processing phase---and decompression upon reception---a similarly resource consuming operation.
Moreover, if multiple nodes transfer similar data chunks across the network hop (\eg, a given pair of switches), each node effectively wastes resources by executing similar steps.
This paper proposes \sys, an approach to design and implement (de)compression at line speed leveraging the \emph{Tofino} hardware platform which is programmable using the \pfoursixteen language.
We report on lessons learned while building the system and show throughput, latency and compression measurements on synthetic and real-world traces, showcasing the benefits and trade-offs of our design.
 	\end{abstract}

	\maketitle

    \acresetall
	\section{Introduction}

Programmable switches allow network engineers to have fine-grain control of the packet flow as well as execute custom operations on those packets, effectively leading to \emph{in-network} processing.
This new paradigm opened the path to offload cumbersome and resource-intensive computations directly into the network data plane.
It has been shown to lead to significantly reduced resource usage at end nodes and increase in network throughput~\cite{10.1145/3317550.3321439}, despite the special care required to handle novel families of threats~\cite{10.1145/3317550.3321436}.
In-network computing has been applied to several domains, including packet aggregation~\cite{10.1145/3152434.3152461}, databases~\cite{lerner2019case}, machine-learning acceleration~\cite{10.1145/3229591.3229594}, data analytics~\cite{234837}, network telemetry~\cite{deepinsight}, and even consensus protocols~\cite{9095258}.
This paradigm can also result in significant energy savings~\cite{10.1145/3302424.3303979} and even offer novel ways to defend against byzantine behaviors~\cite{9013772}.
  
The P4 language~\cite{bosshart2014p4} is becoming the \emph{de facto} standard supported by switches with programmable data planes.
One reason for its popularity is its programming flexibility and the ability to easily verify or test P4 programs~\cite{10.1145/3326285.3329040,10.1145/3185467.3185499}.
Its adoption is quickly rising, including software-defined network (virtualized) settings~\cite{shahbaz2016pisces,10.1145/2999572.2999607} and in hardware to synthesize low-level descriptions for \ac{FPGA} boards~\cite{7544769}.
Lately, network appliance vendors have started offering native support of P4 with comprehensive toolchains, thence paving the way to novel and easy-to-maintain network-related applications which are beyond the limits of \ac{FPGA} implementations.

This paper presents \sys, the first in-network data (de)compressor operating at line-speed.
\sys leverages a novel data deduplication technique: \ac{GD}~\cite{Vestergaard2019-a}, detailed in \autoref{sec:background-gd}. 
Due to the paradigm change imposed by the target hardware, we revise the processing workflow of \ac{GD} (\autoref{sec:approach}).
\sys exploits \pfoursixteen on \ac{TNA} to deploy our implementation on the Barefoot Tofino programmable \ac{ASIC}~\cite{tofino}.

The contributions of this paper are as follows.
We introduce some theoretical background on \ac{GD} in \autoref{sec:background-gd}. 
We survey related work in \autoref{sec:rw}.
Then, we describe how to implement \ac{GD} on Tofino switches, a non-straightforward endeavor (\autoref{sec:approach} and \autoref{sec:implementation}).
We report on several implementation caveats and the lessons learned while doing so (\autoref{sec:lessons-learned}).
Finally, we show that both static and dynamic learning versions of our approach can achieve line rate speed (\ie, \SI{100}{\giga\bit\per\second} on our hardware) for compression and decompression using synthetic and real-world packet traces, thus offloading end nodes from the heavy (de)compression burden (\autoref{sec:evaluation}).
We conclude and hint at future work in \autoref{sec:conclusion}.

\section{Background}

This section provides an overview of our compression algorithm and the coding techniques used for its implementation.

\smallskip\noindent\textbf{\Acf{GD}.}
\label{sec:background-gd}
\Ac{GD} is a generalization of the concept of data deduplication~\cite{Vestergaard2019-a}, where the system first applies a transformation function on a data chunk to split it into a pair of values: a basis and a deviation.
Then, the system proceeds to deduplicate that basis against previously received ones.
The (small) deviation is kept to be able to invert the process upon reading the data.

For example, a transformation function like the one described in \autoref{sec:background-hamming} would map chunks {\small $\{\texttt{0000000}$, $\texttt{0000001}$, $\texttt{0000010}$, $\texttt{0000100}$, $\texttt{0001000}$, $\texttt{0010000}$, $\texttt{0100000}$, $\texttt{1000000}\}$} to a single basis {\small $\texttt{0000}$} and a deviation of $3$ bits, indicating what bit is set (if any).
The same code would map chunks {\small $\{\texttt{1111111}$, $\texttt{1111110}$, $\texttt{1111101}$, $\texttt{1111011}$, $\texttt{1110111}$, $\texttt{1101111}$, $\texttt{1011111}$, $\texttt{0111111}\}$} to basis {\small $\texttt{1111}$}.
In general, the longer the chunk's length (in bits), the more chunks are mapped to the same basis.
The result is that thousands or even millions of chunks can be mapped to the same basis, which increases the potential for compression.

Note that a \emph{dictionary} or registry built based on the most commonly used bases instead of the chunks would be able to represent more (or at least the same number of) chunks, \ie, it allows for a more efficient dictionary to be built.
If we used the above example, a 42-bit sequence {\small $\texttt{|0000000|}$ $\texttt{1111111|}$$\texttt{0100000|}$$\texttt{1111011|}$$\texttt{1000000|}$$\texttt{1011111|}$} has six different 7-bit data chunks, while it can be represented with only two bases.
Two bases in a dictionary could use an ID of only one bit to identify each base.
Thus, the data can be represented with a dictionary of eight bits containing the two bases and a sequence of one-bit basis ID and 3-bit deviations: {\small $\texttt{0|000|1|000|0|111|1|100|0|101|1|110}$} of only 24 bits.
The values for the deviations for this example are described in more detail in \autoref{tab:hamming-syndrom-table}.

Although originally considered for large scale data storage~\cite{Nielsen2019,Talasila2019}, \ac{GD} has been adapted to multi-source data compression protocols~\cite{Gttel2020HermesEE} and file compression for time-series data~\cite{Vestergaard2019-c,Vestergaard2019-b}.
This has resulted in lightweight, online compression mechanisms suitable to the \ac{IoT} an file compressors with excellent random access properties. 

In this paper, we consider Hamming codes as the core component of the transformation function for \ac{GD}, as in~\cite{Vestergaard2019-a,Gttel2020HermesEE}.
The main motivation is that they can be implemented using the \acp{CRC} built-in functionality of Tofino switches for fast execution, as discussed below.
We study cases where the dictionary is preloaded in the switches, or where the sender and receiver dynamically learn it.

\smallskip\noindent\textbf{\Acf{CRC}.}
\label{sec:background-crc}
Let us consider a block of data $B$ with $n$~bits, where $B$ can be expressed as a polynomial $B(x) = b_0 x^0 + b_1 x^1 + ... + b_{n-1} x^{n-1}$ and $b_i$ is the $i$-th bit of $B$.
We consider $b_{n-1}$ as the \ac{MSB} of $B$.
Computing an $m$-bit long \ac{CRC} requires computing the long division of $B(x)$ with a generator polynomial $g(x) = g_0 x^0 + g_1 x^1 + ... + g_{m} x^{m}.$\footnote{We will use $+$ and $\oplus$ interchangeably to represent a XOR operation.}
The residue of this division is the \ac{CRC} value.
One of the properties of \acp{CRC} is that $\ac{CRC}(A \oplus B ) = \ac{CRC}(A) \oplus \ac{CRC}(B)$.
This means that pre-computing all $n$-bit sequences with a single non-zero bit allows us to compute any sequence of $n$ bits by XORing the appropriate \acp{CRC} sequences.
More explicitly, $\ac{CRC}(B) = b_{n-1} \ac{CRC} (1000...0) \oplus b_{n-2} \ac{CRC} (0100...0) \oplus ... b_{1} \ac{CRC} (00...010) \oplus b_{0} \ac{CRC} (00...001)$, which can be represented in matrix form as $ \ac{CRC}(B) = B \vn H^T $.

\smallskip\noindent\textbf{Hamming codes.}
\label{sec:background-hamming}
Hamming codes are block codes that convert $k$~bit messages into $n$~bit messages by adding $m$ parity bits~\cite{blahut_2003}.
More specifically, $n = 2^m -1$ bits and $k = n - m = 2^m -m - 1$ bits.
They are generated using a generator matrix $\vn G$ as follows:
\begin{equation*}
\vn G = \begin{bmatrix}
g_{0,0} & g_{0,1} & \cdots & g_{0,n-1}\\
g_{1,0} & g_{1,1} & \cdots & g_{1,n-1}\\
\vdots & \vdots & \ddots & \vdots 	  \\
g_{k-1,0} & g_{k-1,1} & \cdots & g_{k-1,n-1}
\end{bmatrix}
\end{equation*}
where the size of the matrix is $k \times n$.

Such a code can be transformed to \emph{systematic} form, where the original message is embedded directly into the codeword and the parity bits are clearly separated from it as $\vn G = \begin{bmatrix}\vn I_{k} & \vn P \end{bmatrix}$, which is achieved by performing row operations to eliminate components.
$\vn I_{k}$ indicates an identity matrix of size $k \times k$.

For our work, we will consider a case where the order of parity and identity matrices are shifted, \ie, $\vn G_s =\begin{bmatrix}\vn P & \vn I_{k}\end{bmatrix}$, as it matches the output of \ac{CRC} functions.
The \textit{parity-check matrix}, $\vn H $ of size $m \times n$ is given by $\vn H = \begin{bmatrix} \vn I_{n-k} & \vn P^T\end{bmatrix}$, which is actually the same as for computing the $\ac{CRC}$ with the same generator polynomial~\cite{blahut_2003}.
A message $\vn u$  ($1 \times k$) is encoded into the codeword $\vn c$ ($1
\times n$) by $\vn c = \vn u\vn G_s$.

For decoding, the received sequence is $\vn B = \vn c + \vn e$, where $\vn e$ is the error pattern.
An all-zero $\vn e$ means there are no errors.
The matrix $\vn H$ can be used to detect whether any errors occurred by calculating the \emph{syndrome} vector $\vn s = \vn B \vn H^T = (\vn c + \vn e) \vn H^T = \vn e \vn H^T$, due to the fact that $\vn G_s \vn H^T = \vn 0$ (by construction of $H$ and $G_s$).
A look up table can be constructed that maps the different $1$-bit error patterns to a corresponding $\vn s$ value.

\smallskip\noindent\textbf{Connection of hamming codes to \acp{CRC}.}
\Iac{CRC}-m using a generator polynomial that is suitable to a Hamming code $(n,k)$ can help compute the syndrome and parity bits in the decoding and encoding steps.
More specifically, as long as
\begin{enumerate*}
\item the generator polynomial for a Hamming code is used as parameter for the \ac{CRC}-m generator polynomial \item the size of the input data to be computed by the \ac{CRC}-m is $n = 2^m-1$~bits as expected by a standard Hamming code, then the computed \ac{CRC}-m value will be equivalent to the syndrome computed for a $(n,k) = (2^m-1,2^m-m-1)$ Hamming code
\end{enumerate*}.
\autoref{tab:hamming-generator-poly-table} provides generator polynomials for Hamming codes and the expected input parameter for the \ac{CRC}-m module in Tofino chipsets.

\begin{table}[t]
\caption{Generator polynomials for Hamming codes and parameters for \iac{CRC}-m.}
  \small
\label{tab:hamming-generator-poly-table}
 \resizebox{\linewidth}{!}{\begin{tabular}{lll}
\toprule
  Code                &   Generator polynomial      & Parameter for \ac{CRC}-m  \\
\midrule
  $(7,4)$             &  $x^3+x+1$                  &      $0x3$            \\
  $(15,11)$           &  $x^4+x+1$                  &      $0x3$            \\
  $(31,26)$           &  $x^5+x^2+1$                &      $0x05$           \\
  $(31,26)$           &  $x^5+x^4+x^2+x+1$          &      $0x17$           \\
  $(63,57)$           &  $x^6+x+1$                  &      $0x03$           \\
  $(127,120)$         &  $x^7+x^3+1$                &      $0x09$           \\
  $(255,247)$         &  $x^8+x^4+x^3 +x^2+1$       &      $0x1D$            \\
  $(511,502)$         &  $x^9+x^4+1$                &      $0x00D$           \\
  $(511,502)$         &  $x^9+x^8+x^7+x^6+x^5+x+1$  &      $0x0F3$            \\
  $(1023,1013)$       &  $x^{10}+x^3+1$             &      $0x009$            \\
  $(2047,2036)$       &  $x^{11}+x^2+1$             &      $0x005$            \\
  $(4095,4083)$       &  $x^{12}+x^6+x^4+x+1$       &      $0x053$            \\
  $(8191,8178)$       &  $x^{13}+x^4+x^3+x+1$       &      $0x01B$            \\
  $(16383,16369)$     &  $x^{14}+x^8+x^6+x+1$       &      $0x143$            \\
  $(32767,32752)$     &  $x^{15}+x+1$               &      $0x003$            \\
\bottomrule
\end{tabular}
}
\end{table}
     \section{Related Work}
\label{sec:rw}
Programmable network switches have sparked a lot of interest in academia and industry.
In particular, the P4 language allows implementing novel tools, \eg, real-time network telemetry and analytics~\cite{tofigh,lapukhov}, and  accelerating existing applications, \eg, stream processing~\cite{10.1145/3185467.3185494}.
One of the reasons for P4's current momentum is that it allows easy verification of in-network programs~\cite{verification-p4-conext18,10.1145/3341216.3342206}.
We will exploit this opportunity in future work.

P4 on Tofino has been successfully used to implement NetEC~\cite{in-network-erasure-coding}, an efficient in-network \ac{EC}.
Data streams arrive from different switch ports, aggregated using Reed-Solomon \ac{EC}~\cite{reed1960polynomial}, and leave the switch from yet another port in aggregated fault-tolerant form.
Both \sys and NetEC offload heavy-duty processing from end-node CPUs: packet compression and packet encoding, respectively.
While NetEC mainly operates over packet payloads, P4-enabled switches can be used to deploy more generic forms of aggregations and disaggregation, \eg, over several header frames for fixed-size~\cite{wang2019high} or variable-size~\cite{aggregating-packets} payloads, while still achieving line-rate throughput.
In a nutshell, small payloads are stored in the switch's register array and, upon a new small packet arrival, they are recirculated via the ingress parser until the \ac{MTU} is reached, \ie until the aggregation is complete.

Chen et al.~\cite{implementing-aes} presented a non-trivial design and implementation for an in-network symmetric encryption scheme (AES) for different payload sizes, where packets go through the switch pipeline multiple times to complete the 10 rounds used for AES-128.
In contrast, \sys does not need packet recirculation as \ac{GD} can be implemented in a single round, which translates in higher speed and lower delay.

Although there exist switches performing on-the-fly compression of data streams~\cite{brocade6520,XA-200K,aruba}, these operate over packets at layer 3 or above. 
Our approach can support a wider set of network and transport protocols by operating at layer 2.

Since \sys leverages \ac{GD}, it performs particularly well under very limited memory constraints~\cite{yazdani2020memory}.
For instance, DEFLATE~\cite{deutsch1996rfc1951} is a well-known standard compression technique, which requires a minimum of \SI{3}{\kilo\byte} to compress data.

\ac{IoT} applications normally deal with small data chunks and compression per chunk due to memory limitations.
However, standard compression techniques are known to perform poorly over small data~\cite{yazdani2019protocols}.
\ac{GD} performs particularly well for small data chunks, which makes \sys suitable for a wide range of chunk sizes.
In many application scenarios, such as live video streaming, freshness of information plays a key role.
Standard compression techniques,\ie DEFLATE, require enough data to provide satisfactory compression, negatively affecting the freshness of information, as collection of such data takes time.
\sys is perfectly suitable for applications with timeliness restrictions, as \ac{GD} provides significant improvements in terms of freshness of information compared to state-of-the-art techniques~\cite{yazdani2020age}.
 	\section{Approach}
\label{sec:approach}

\begin{figure}[t!]
  \centering
  \includegraphics[width=\linewidth]{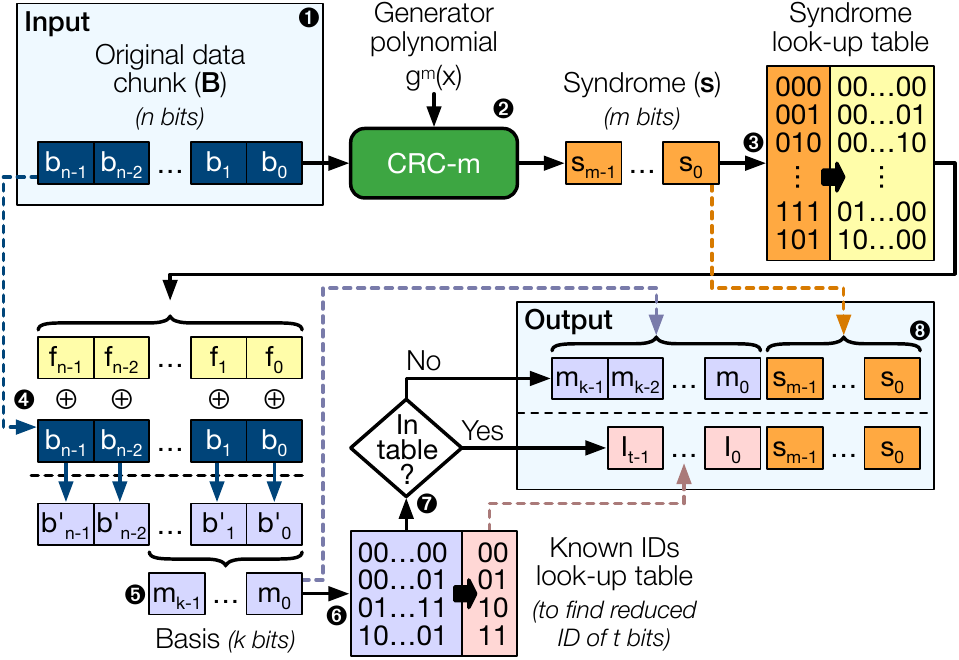}
  \caption{\acs{GD} encoding workflow on Tofino.}
  \label{fig:gdd-enc}
\end{figure}

Our approach uses the \ac{GD} algorithm presented in \autoref{sec:background-gd} to implement compression of network packets in a way that is implementable on a readily-available programmable switch.
The \emph{encoding} workflow is shown in \autoref{fig:gdd-enc}.
Let us consider a data payload $B$ of size $n$, part of an incoming network packet~\ding{202}.
The first action to perform is to compute the syndrome~$s$ using a Hamming decoder (mapped to an equivalent \ac{CRC})~\ding{203}.
The syndrome tells us which bit in the original data needs to be flipped.
We achieve this by having a table~\ding{204} that matches to the correct mask~$f$ to be XORed to the original data~$B$~\ding{205} according to $s$.
The result of this XOR operation~$b'$ is truncated to the rightmost $k$ bits to form a basis~$(m_{k-1},...,m_{0})$~\ding{206}.
As several data chunks share the same basis, we take this opportunity to replace them with shorter identifiers~$I$ should the basis been seen before~\ding{207},\ding{208}.
In order for the recipient to be able to find $B$ again, we  attach the syndrome~$s$ to the compressed packet~\ding{209}.

\begin{figure}[b!]
	\centering
	\includegraphics[width=\linewidth]{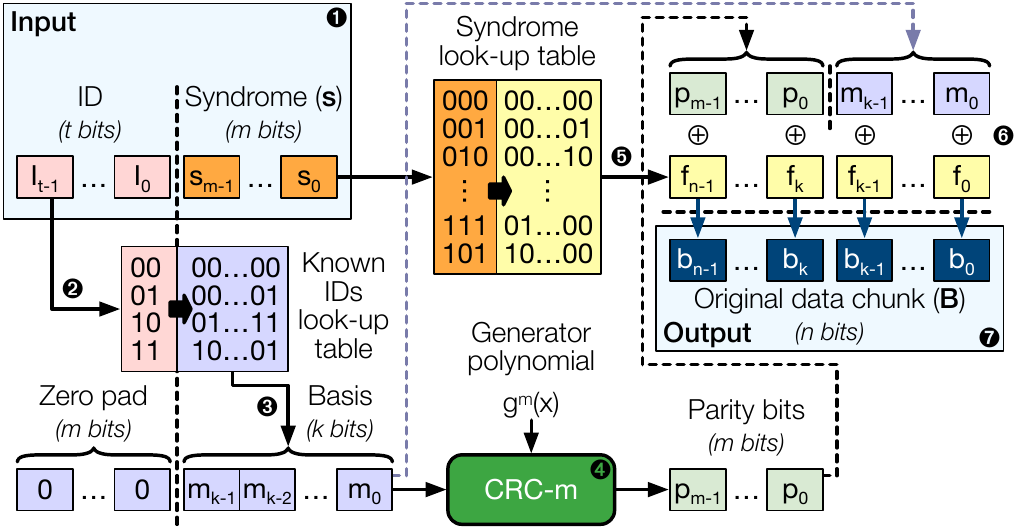}
	\caption{\acs{GD} decoding workflow on Tofino.}
	\label{fig:gdd-dec}
\end{figure}

The decoding workflow is shown in \autoref{fig:gdd-dec}.
We consider that a packet containing a compressed basis has arrived~\ding{202}.
For an uncompressed basis, the workflow instead starts at step~\ding{204}.
On the recipient's side, there must exist a table that maps short identifiers~$I$ to basis that is synchronized with its counterpart in the encoder~\ding{203}.
We zero-pad the recovered basis~\ding{204} and then feed it to the same \ac{CRC} generator as on the encoder~\ding{205} to get $p$, hence restoring the bits that we truncated in the encoder.
In parallel, we use the same syndrome look-up table as in the encoder to identify which bit needs to be flipped according to~$s$.
This gives us the appropriate mask~$f$ to be XORed~\ding{206}.
We apply the mask to flip the right bit out of the concatenation of $p$ and the basis~\ding{207} to successfully restore the original data~$B$~\ding{208}.
 	\section{Implementation}
\label{sec:implementation}
Our implementation uses the special capabilities of the Tofino chip from Barefoot Networks.
We use the \pfoursixteen language, with Tofino-specific proprietary extensions, \ie, \ac{TNA}, and Barefoot \ac{SDE} version~9.2.0.
The control plane is implemented using Python and \ac{BfRt}.
We favored simplicity whenever possible.
In this mindset, we chose to develop everything from scratch, which also reduces the likelihood of resource allocation issues.
We settled on Ethernet-based framing to provide compatibility with regular Ethernet network cards and guarantee inter-connectivity.

We define three different types of packets:
\begin{enumerate*}
	\item\label{packet-regular} regular, yet unprocessed packets
	\item\label{packet-dedup-basis} processed, but uncompressed packets, comprised of a \emph{basis} and a \emph{syndrome}
	\item\label{packet-dedup-id} processed and compressed packets, which replace the \emph{basis} with a shorter identifier
\end{enumerate*}.
Packet type \ref{packet-regular} can be any Ethernet packet that arrives in the switch.
Our implementation takes such packets as input, and then transforms them into types \ref{packet-dedup-basis} or \ref{packet-dedup-id}.
The latter type can only be produced when there exists a basis-ID mapping for the basis that the original type \ref{packet-regular} packet maps to.
Such mappings are initialized once an unknown basis is computed.
They remain valid for as long as possible.
\Iac{LRU} cache eviction policy decides when to recycle an identifier to map to a different basis.
More details about this stateful part of our implementation are provided in later paragraphs.

\begin{table}
\caption{Hamming code $(7,4)$ and CRC-3 equivalence}
	\label{tab:hamming-crc}
	\small
\subfloat[Hamming code $(7,4)$\\syndromes]{\begin{tabular}{ccc}
			\toprule
			Error    & Bit sequence     &   Syndrome    \\
			\midrule
			$0$         & $(0000001)$      &   $(001)$        \\
			$1$         & $(0000010)$      &   $(010)$        \\
			$2$         & $(0000100)$      &   $(100)$        \\
			$3$         & $(0001000)$      &   $(011)$        \\
			$4$         & $(0010000)$      &   $(110)$        \\
			$5$         & $(0100000)$      &   $(111)$        \\
			$6$         & $(1000000)$      &   $(101)$        \\
			\bottomrule
		\end{tabular}\label{tab:hamming-syndrom-table}
	}
	\hfill
	\subfloat[CRC-3 of bit sequences\\with one non-zero bit]{\begin{tabular}{ccc}
			\toprule
			Poly.    & Bit sequence     &   \texttt{CRC-3}    \\
			\midrule
			$x^0$         & $(0000001)$      &   $(001)$        \\
			$x^1$         & $(0000010)$      &   $(010)$        \\
			$x^2$         & $(0000100)$      &   $(100)$        \\
			$x^3$         & $(0001000)$      &   $(011)$        \\
			$x^4$         & $(0010000)$      &   $(110)$        \\
			$x^5$         & $(0100000)$      &   $(111)$        \\
			$x^6$         & $(1000000)$      &   $(101)$        \\
			\bottomrule
		\end{tabular}\label{tab:crc3-table}
	}
\end{table}

The foundation of our implementation is the \ac{GD} algorithm as described in \autoref{sec:background-gd}.
Notably, \ac{GD} uses Hamming codes to work as they are equivalent to some particular \acp{CRC}.
\autoref{tab:hamming-crc} shows an example of the equivalence between the $(7,4)$ Hamming code and CRC-3 with a carefully chosen generator polynomial.
The Tofino platform offers a native component to compute \acp{CRC}, as such codes are commonly found in many network protocols.
We extensively rely on this component to efficiently implement the key steps of the \ac{GD} algorithm on Tofino, namely the computation of \emph{syndromes}.

The next part of the workflow requires flipping one bit in the input data according to the syndrome that we obtained in the previous step.
We use a P4 table with constant entries that are pre-computed using a short C++ program making use of Boost CRC library~\cite{boost-crc}.
The entry that matches the syndrome is XORed to the data, hence flipping the appropriate bit of the sequence.
This transformation creates the \emph{basis}.

In order to allow for the compression of packets, we need to replace \emph{$\text{syndrome}+\text{basis}$} couples with shorter identifiers.
One possible way to select an identifier for a given basis would be to use cryptographic hashes, which cannot be computed in one pass on our programmable switch.
The alternative is to use arbitrary identifiers~(IDs).
We involve the control plane to manage the pool of identifiers.
Unknown bases are sent up by means of \emph{digests}, as provided by \pfoursixteen/\ac{TNA}.
Recording a new basis-ID mapping is done in two phases; first, the control plane chooses an identifier to assign to the basis.
When there are unused identifiers, the control plane selects the least recently used one.
Should all identifiers be in use, \iac{LRU} policy is applied to evict and recycle an identifier to accommodate the most recent basis.
Setting \iac{TTL} that is automatically decreased as time elapses on a particular table entry is a feature that is provided by \ac{TNA}.
With the identifier now selected, the control plane first sets the reverse mapping (ID-basis) in the destination switch to make sure that compressed packets can always be uncompressed.
The control plane can finally add a corresponding entry in the source switch, to map the newly-discovered basis to the chosen identifier.
Subsequent packets sharing the same basis are then transmitted in a shorter \emph{$\text{syndrome}+\text{identifier}$} form.

Last, we add \emph{counters} to our program to provide easily-accessible statistics of the inner-workings of \sys.
In particular, packets are classified according to how they are transformed (\eg, raw packet to $\text{syndrome}+\text{basis}$, $\text{syndrome}+\text{identifier}$ to raw packet, \etc).
 	\section{Lessons learned}
\label{sec:lessons-learned}
We report on several lessons that we learned during the development process of \sys.
In particular, we highlight some potential intricacies that we encountered when developing \sys with \pfoursixteen/\ac{TNA} for the Barefoot Tofino platform.
While \pfoursixteen is a flexible language, it is mainly designed for efficient processing of network packet headers and routing.
Similarly, the hardware design of the Tofino chip is tailored to the same goals.
Our quite unorthodox use of the combination of \pfoursixteen and the underlying hardware is far from traditional packet routing.
As a consequence, our implementation had to be adapted to circumvent some resulting limitations.

\emph{Header} declarations in \pfoursixteen must be aligned on byte boundaries.
However, the Hamming codes that we use are never aligned on bytes.
While \pfoursixteen itself treats data at bit-level granularity, the hardware---and the compiler by extension---are optimized for byte-aligned data.
A side effect is that we have to introduce padding bits in our \pfoursixteen program when selecting particular sizes that are not byte-aligned for the program to compile.
In turn, this introduces a loss of goodput when these sizes are selected, limiting the range of interesting parameter values to only a few.

In our original design, we placed as much of the code as possible in the data plane.
This was made possible by leveraging \emph{registers}, \ie, user-accessible memory in the Tofino data plane pipeline.
Doing so allowed us to achieve line-rate performance and virtually instantaneous learning of new basis-ID pairs.
However, as every piece of code running in the data plane must be able to execute in constant time, many algorithms are out-of-reach, \eg, algorithms that need a complete view over an array of registers.
Therefore, we settled on storing basis-ID pairs in regular match-action tables and manage them with the control plane, which is implemented in a regular programming language (\ie, Python in our case).
This allows us to use features such as \emph{digests} and \emph{per-table-entry \acp{TTL}} that the \ac{TNA} framework provides, conveniently letting us implement \iac{LRU} cache for basis-ID pairs.
We can also send updates regarding ID-basis pairs to other \sys instances out-of-band.
The drawback of this approach is that updates take a longer time to effectively apply (see \autoref{sec:eval:learning}).

Finally, we recognize that the Barefoot Tofino platform is flexible in many aspects.
One such aspect is the unified ingress-egress pipeline.
It is possible to artificially extend the pipeline without recirculating packets by shifting parts of the processing (\ie, \ac{GD} decoding) to the egress.
This provides space for more table entries, as the compiler can place egress tables across stages that the ingress pipeline underuses.
     \section{Evaluation}\label{sec:evaluation}

Our setup leverages an Edgecore Wedge100BF-32X programmable switch~\cite{wedge-datasheet}.
It is connected at \SI{100}{\giga\bit\per\second} to two Dell PowerEdge R7515 servers through Mellanox ConnectX-5 network cards interfaced with PCI~Express 3.0 x16.
Each server has an AMD EPYC 7302P processor, \SI{32}{\gibi\byte} of RAM and runs Ubuntu 20.04.1.
We tune each server with the \texttt{mlnx\_tune} utility in \texttt{HIGH\_THROUGHPUT} mode.
Unless stated otherwise, each measurement is repeated 10 times, and we show the average and the \SI{95}{\percent} confidence interval.

\smallskip\noindent\textbf{Choice of parameters.}
It is possible to set several parameters in \sys.
Three of them pertain to Hamming codes: $k$, $n$ and $m$ (as introduced in \autoref{sec:background-hamming}).
Specifically, the values of $k$ and $n$ are strongly linked to the value of $m$ by formulas, so only $m$ can in fact be freely selected.
Since the Tofino platform has explicit byte-alignment constraints on header fields, every value of $m$ that is not a multiple of \num{8} requires us to include useless padding bits in packets.
We select $m=8$ which is the largest $m$ that is a multiple of \num{8} and that fits hardware constraints.

Additionally, it is possible to independently choose the size of the short identifiers that replace the bases.
This parameter dictates how many bases have to be cached by \sys.
Again, byte-alignment constraints have to be adhered to when choosing this value, such that useless padding bits can be omitted.
We require one additional bit to store the \ac{MSB} of the raw data packet---its size, $k$, is always one below a multiple of \num{8}.
Therefore, this parameter also needs to be one below a multiple of \num{8} to omit useless padding bits.
Akin to our choice of $m=8$, we settle for the largest value that is one below a multiple of \num{8} and fits hardware constraints (especially in terms of resource usage in this case): \SI{15}{\bit}, allowing for $2^{15}=\num{32768}$ cached bases.

\smallskip\noindent\textbf{Dynamic learning.}
\label{sec:eval:learning}
We want to measure the consequence of our decision to move the code responsible for managing bases-ID pairs to the control plane (see \autoref{sec:lessons-learned}).
In the data-plane, every stateful operation appears to be instantaneous and already applies to the next packet in the pipeline.
The main drawback of involving the control plane relates to performance, as it is located away from the path that packets take.

In this experiment, we measure the time between the arrival of an unknown basis in the switch and the moment after which the basis is registered in the compression table, and compressed packets start to be produced.
To do so, we repeatedly send the same data packet as fast of possible from one server to another.
We capture packets on the destination server and measure the amount of time it takes between the arrival of the first packet of type~\ref{packet-dedup-basis} and the arrival of the first packet of type~\ref{packet-dedup-id} (see \autoref{sec:implementation} for the definitions of types).
Our results indicate that it takes \SI{1.77 +- 0.08}{\milli\second} for \sys to record and apply a new basis-ID pair.
During that time, packets that share the same basis stay uncompressed.
This loss in compression efficiency is measured next.

\smallskip\noindent\textbf{Compression.}
\label{sec:eval-compression}
The goal of this experiment is to assess the compression ratio that can be obtained by using \sys.
We use two datasets, a synthetic and a real-world one.
We engineered the synthetic dataset to be behaviorally close to typical readouts from a sensor.
We generate \num{3124000} chunks of \SI{256}{\bit} (matching the parameters we chose), which are then converted to a \emph{pcap} trace of Ethernet packets containing the chunks as payload.

The real-world dataset consists in a day of DNS queries at a \num{4000} users university campus \cite{dns-dataset}.
To obtain the trace that we replay, we apply filters to the downloaded files to only keep queries of \SI{34}{\byte} going to the main DNS resolver of the campus, excluding the DNS transaction identifer which is a random number.

We replay these traces to our switch and monitor which action \sys undertakes with the payload of each packet.
We then deduce the payload size, as each action produces a packet type of a fixed size.
\begin{figure}[!t]
    \raggedright
    \includegraphics{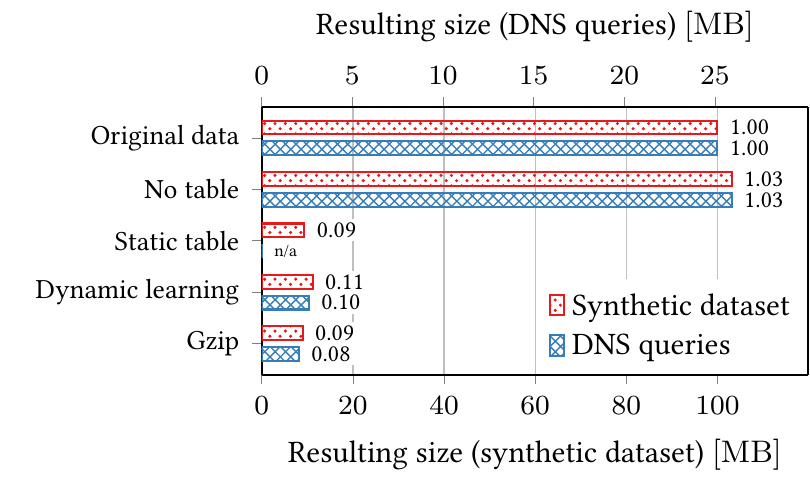}
    \caption{Resulting payload size after traffic is processed with Gzip and \sys, without, with static-, and with dynamically learned- compression table mappings.}
    \label{fig:eval-compression}
\end{figure}
The sum of all original chunks represents the baseline.
In \autoref{fig:eval-compression}, the bars represent the total size of the payloads of all packets after they are transformed by \sys.
We also show the numerical ratio to the baseline next to each bar.

We measure three cases:
\begin{enumerate*}
    \item\label{enum:notable} \emph{no table}: the compression table stays empty
    \item\label{enum:static} \emph{static table}: we pre-compute the basis of each payload and add a corresponding mapping in the compression table before we start the experiment
    \item\label{enum:dynamic} \emph{dynamic learning}: we start the experiment with an empty compression table, which is then automatically filled by \sys when unknown bases are encountered
    \item\label{enum:gzip} \emph{Gzip}: we extract all payloads in a regular file that we compress with the \texttt{gzip} compression tool
\end{enumerate*}.
In theory, case \ref{enum:notable} should be equal to the baseline as applying \ac{GD} does not introduce additional bits.
The \SI{3}{\percent} overhead is due to padding bits which are necessary to guarantee container alignment on the Tofino platform.
We reckon that 8 such padding bits could be eliminated by an expert \pfoursixteen/\ac{TNA} programmer.
Case (\ref{enum:static}) represents the idealistic scenario where the basis of every packet payload is known beforehand.
Case (\ref{enum:dynamic}) represents a real scenario where the traffic is unknown to the switch prior to its arrival.

We see that \sys correctly learns and stores bases in its compression table, providing savings of \SI{89}{\percent} in terms of bytes transmitted in the synthetic scenario, and up to \SI{90}{\percent} in the DNS dataset.
The compression ratio of \sys compares well with an off-the-shelf compression utility (\circa \SI{20}{\percent} difference), \texttt{gzip}, which uses an algorithm (DEFLATE) that doubtlessly cannot be implemented on our hardware P4 target due to its unbounded execution time.

The delta between cases (\ref{enum:static}) and (\ref{enum:dynamic}) in the synthetic scenario is first due to the fact that one packet with a payload mapping to each basis must be transmitted without compression to let the recipient know about it.
Second, a couple milliseconds are needed between the time a packet with an unknown basis arrives in the switch, and the mapping to its basis becomes effective for subsequent packets (as shown above), meaning that additional packets sharing the same basis stay uncompressed during this processing delay.

\begin{figure}[!t]
    \centering
    \includegraphics{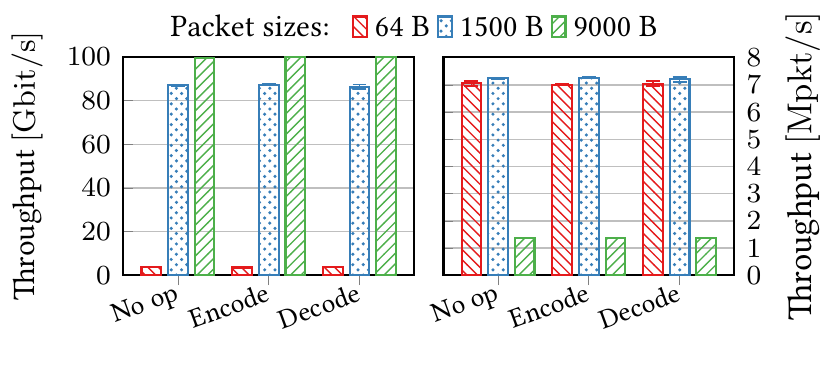}
    \caption{Observed network throughput in \si{\giga\bit\per\second} (left) and \si{\mega\packet\per\second}\si{{packet}\per\second} (right) with the switch performing various operations on Ethernet frames of different sizes.}
    \label{fig:eval-tput}
\end{figure}

\smallskip\noindent\textbf{Raw performance.}
Finally, we measure the raw performance of \sys using \texttt{raw\_ethernet\_} utilities provided by Mellanox.
We start by measuring the raw Ethernet throughput between 2 machines through the programmable switch.
We transfer Ethernet frames of 3 common sizes for 10 seconds: the minimum frame size of \SI{64}{\byte}, the standard \SI{1500}{\byte}, as well as \emph{jumbo} frames of \SI{9}{\kilo\byte}.
The first scenario (``no op'') acts as the baseline, with the switch acting as a regular Ethernet switch.
We then repeat the same measurements with the switch performing either the encoding or the decoding phase of \sys.
\autoref{fig:eval-tput} shows our results.
We notice that the claims put forwards by the vendor of our programmable switch are kept; namely that any \pfoursixteen program that successfully compiles for the Tofino platform performs at line speed, so long as it does not make use of recirculation, packet duplication, \etc.
The figures for \num{64} and \SI{1500}{\byte} packets are bottlenecked at around \SI{7}{\mega\packet\per\second} by the server generating the traffic.
In theory, the full line rate of \SI{100}{\giga\bit\per\second} can also be reached with these smaller packets, as per the \SI{4.7}{\giga\packet\per\second} figure quoted in the datasheet of the switch~\cite{wedge-datasheet}.

\begin{figure}[!t]
    \raggedright
    \includegraphics{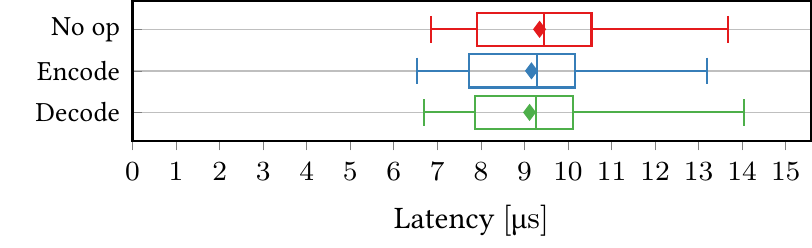}
\caption{Observed end-to-end latency with the programmable switch performing various operations.}
    \label{fig:eval-latency}
\end{figure}

Subsequently, we evaluate the latency by having one server sending packets to itself via the programmable switch.
We then measure the \ac{RTT}.
Results are shown in \autoref{fig:eval-latency} and point towards the same findings than our throughput measurements: the addition of \sys to the \pfoursixteen program has no noticeable effect on raw performance.
     \acresetall
	\section{Conclusion and future work}
\label{sec:conclusion}
This paper introduces \sys, an approach for in-network data compression that can operate at line rate with minimal delay on (de)compression.
We adapt the concept of \ac{GD} to provide an efficient implementation and a mapping of the data that can significantly boost the benefits of limited dictionaries implemented in the switches.
We rely on transformations based on Hamming codes, which can be implemented efficiently with the switch's \ac{CRC} components.
Our implementation shows that (de)compression at line rate (\SI{100}{\giga\bit\per\second} in our switch) is possible and that both static tables (for the dictionary) and dynamic learning can be implemented without compromising throughput or compression gains.

The use of the \ac{CRC} module in Tofino switches opens the door to computation of more complex transformations, \eg, BCH codes, by using different generator polynomial parameters~\cite{blahut_2003}.
These allow for more chunks to be mapped to each basis, albeit at the cost of a larger deviation in bits.
Future work will explore alternative transformations with efficient implementations, \eg, bit-swapping~\cite{Vestergaard2019-c}, and dynamic approaches to select a wider range of parameters.

\begin{acks} 
	We are thankful to Vladimir Gurevich (Barefoot Networks) for his insightful feedback.
	This work was partially financed by the SCALE-IoT Project (Grant No. 7026-00042B) granted by the Independent Research Fund Denmark, by the Aarhus Universitets Forskningsfond (AUFF) Starting Grant Project AUFF- 2017-FLS-7-1, and Aarhus University’s DIGIT Centre.
\end{acks}

	\bibliographystyle{ACM-Reference-Format}
	\bibliography{references}
\end{document}